\documentclass{aa}
\usepackage{psfig,natbib}
\bibpunct{(}{)}{;}{a}{}{,} 

\defcitealias{iso_nature}{Mor99}
\defcitealias{polomski_99}{Pol99}
\defcitealias{pantin_timmi}{PL00}

\begin{document}
\headnote{Letter to the Editor}
\title{Discovery of a double ring in the core of
  $\eta$~Carinae
  \thanks{Based on observations taken at the European Southern
    Observatory, La Silla, Chile}}

\author{
  S. Hony\inst{1},
  C. Dominik\inst{1},
  L.B.F.M. Waters\inst{1,2},
  V. Icke\inst{3},
  G. Mellema\inst{3},
  R. van Boekel\inst{4,1},
  A. de Koter\inst{1},
  P.M. Morris\inst{5},
  M. Barlow\inst{6},
  P. Cox\inst{7},
  H.U. K\"aufl\inst{4}}

\authorrunning{Hony et al.}
\titlerunning{A double ring in $\eta$~Car}

\institute{
  Astronomical Institute ``Anton Pannekoek'', University of Amsterdam,
  Kruislaan 403, NL--1098 SJ Amsterdam, The Netherlands
  \and
  Instituut voor Sterrenkunde, Katholieke Universiteit Leuven,
  Celestijnenlaan 200B, B--3001 Heverlee, Belgium
  \and
  Sterrewacht Leiden, Postbus 9513, NL--2300 RA Leiden, The Netherlands
  \and
  European Southern Observatory, Karl-Schwarzschild Strasse 2,
  D--85748 Garching, Germany
  \and
  SIRTF Science Center, Caltech, Mail Code 314-6, 1200 East California
  Boulevard, Pasadena, CA 91125, USA
  \and
  University College London, Department of Physics and Astronomy,
  Gower Street, WC1E 6BT London, U.K.
  \and
  Institut d'Astrophysique Spatiale, B\^at. 121,
  Universit\`e de Paris Sud, F91405 Orsay Cedex, France
  }

\offprints{Sacha Hony (hony@astro.uva.nl)}
\date{Received $<$date$>$; accepted $<$date$>$}

\abstract{We report the discovery of a double ring structure in the
  waist of the nebula surrounding $\eta$~Carinae.  The rings are
  detected in the mid-IR dust continuum at wavelengths of 7.9, 11.9,
  12.9 and 20 $\mu$m. The dust in the rings has a temperature of about
  300~K.  The orientation of the rings is inclined with respect to the
  axis of the homunculus by either 37 or 58 degrees. The central star
  is not in the projected centre of the structure defined by the two
  rings. This geometry is reminiscent of that seen in SN1987A and some
  planetary nebulae. We discuss several possible origins for this
  remarkable geometry and its orientation.  
  }

\maketitle

\section{Introduction}
The star \object{$\eta$~Car} is one of the best studied, but still
poorly understood members of the class of Luminous Blue Variables
\citep{davidson_araa97}.  $\eta$~Car shows a spectacular, strongly
bipolar nebula referred to as the homunculus
\citep[e.g.][]{morse_hst}, that was created during the great eruption
of 1840$-$1860. The homunculus is sometimes described as a hollow
flask expanding with velocities between 400 and $\sim$800~$\rm
km\,s^{-1}$.  A conspicuous equatorial skirt is seen in the Hubble
Space Telescope (HST) images.  Proper motion studies of the
condensations in the skirt show that these were ejected during the
great eruption \citep{morse_bullets}.  There is strong evidence that
$\eta$~Car is a binary star with a period of 5.52 years and orbital
eccentricity of 0.6 \citep{damineli_1996,damineli_2000}.  The X-ray
light curve of $\eta$~Car is consistent with a colliding winds model
\citep{corcoran_xte}, and supports the binary model for $\eta$~Car.

Many studies have focused on infrared imaging of the dusty homunculus
(e.g. \citealt{gehrz_73,smith_98}; \citealt[][hereafter
Pol99]{polomski_99}; \citealt{gehrz_millenium}; \citealt[][hereafter
PL00]{pantin_timmi}).  These studies show that at near-IR wavelengths
the central star dominates while the lobes are faint. Note that
speckle polarimetry of the core of $\eta$~Car indicates that the
central star is \emph{not} viewed directly but is obscured
\citep{falcke_speckle}. At mid-IR wavelengths several equatorial
emission blobs northeast and southwest of the central star are
visible. \emph{These blobs are interpreted as evidence for an
  equatorial torus} (\citetalias{polomski_99}; \citealt[][hereafter
Mor99]{iso_nature}).  \citetalias{pantin_timmi} show that the bipolar
lobes cannot be empty and suggest the presence of a second shell
interior to the optically visible homunculus. HST observations also
point to an inner structure \citep{ishibash_hst}.

The Infrared Space Observatory (ISO) spectrum of $\eta$~Car (taken
Jan. 1996) shows three spectral components: a power law between 2 and
about 10 $\mu$m, arising from the central point source, a
T~$\sim$~250~K component, attributed to the lobes, and a
T~$\sim$~110$-$130~K component, attributed to a cold,
$\sim$~15~M$_{\odot}$ equatorial torus \citepalias{iso_nature}. It was
suggested that the massive equatorial torus was present before the
great eruption and caused the highly bipolar shape of the nebula.

In this \emph{Letter} we discuss new mid-IR images of $\eta$~Car that
resolve the equatorial blobs and show that they have a highly
symmetric ring-like shape with a surprising orientation with respect
to the homunculus.  The images may point to a drastic change in the
orientation of the plane of symmetry of the system some time after the
great eruption.

\section{The Observations}
\begin{figure}[!t]
  \psfig{figure=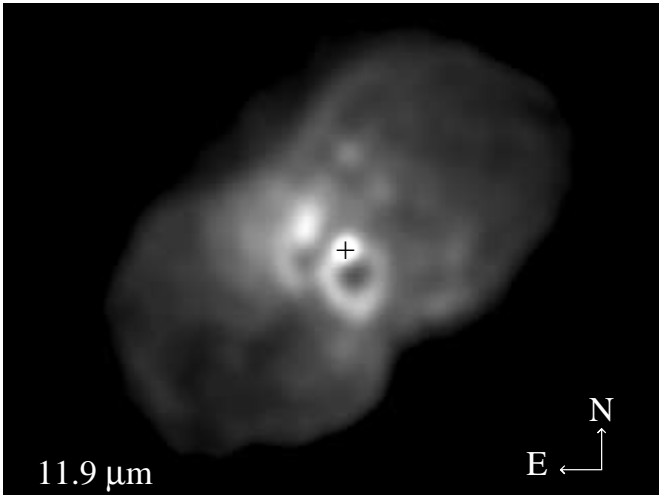,width=8.7cm}
  \caption{TIMMI2 image of $\eta$~Car in the 11.9 $\mu$m narrow band
    filter. We applied a square root function to enhance the
      intensity contrast. The '+' indicates the central source.}
  \label{fig:nebula}
\end{figure}
We observed $\eta$~Car between 3 and 6 March 2001, with the TIMMI2
\citep{timmi2_2000,timmi2_messenger} mid-IR camera attached to the ESO
3.6m telescope at La Silla, Chile.  The camera is equipped with a
320x240 pixel array; we applied a pixel scale of 0.2 arcsec/pixel in
both the N and the Q band.  We used a chop throw of 22 arcsec
north-south and a nod of 30 arcsec east-west.  This allowed for both
the chopped as well as the nodded positions to fall onto the detector.
In the final images, we combined the positive and negative images
resulting from the chopping and nodding positions.  $\eta$~Car was
observed in the M band (4.8 $\mu$m), at several wavelengths in the N
band using narrow-band filters (7.9, 11.9, 12.9 $\mu$m), and in the Q
band (20 $\mu$m).  The observing conditions were variable with a high
humidity and seeing ranging between 0.3 and about 1 arcsec.  Apart
from the images, a number of long-slit N-band spectra covering the
entire nebula were taken, as well as a few Q-band spectra.  In this
paper, we concentrate on the 7.9, 11.9, 12.9 and 20 $\mu$m images.  We
will report on the other images, as well as on the spectroscopy in a
subsequent study.

The images were reduced using a shift-and-add technique, where the
shift between individual frames was determined from a least-squares
comparison between the images. The resulting images were then
deconvolved using an empirically determined point spread function
obtained by observing \object{$\gamma$~Cru} with the same set-up as
$\eta$~Car.  For the Q band image we have no good empirical point
spread function.  However, given that in the Q band TIMMI2 at the 3.6m
is diffraction-limited, we have adopted a Gaussian beam of 1.25 arcsec
to deconvolve the Q band image.

\begin{figure}[!t]
  \psfig{figure=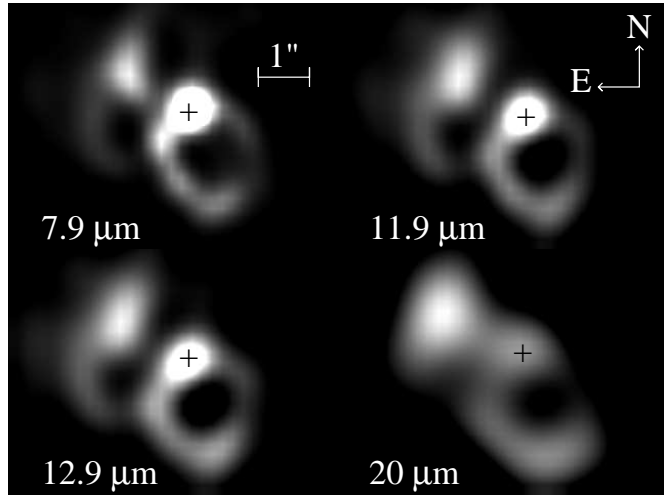,width=8.7cm}
  \caption{Multi wavelength observations of the central part of the
    $\eta$~Car nebula. The double ring structure is traced
      at all four wavelengths. The '+' indicates the central source.}
  \label{fig:rings}
\end{figure}

\begin{figure}[!t]
  \centerline{
  \psfig{figure=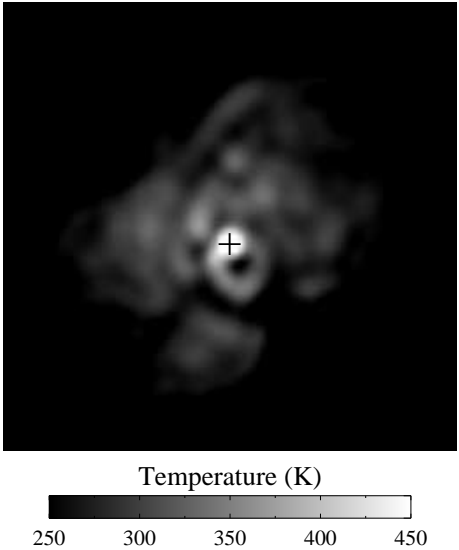,width=6.0cm}}
  \caption{Temperature map as derived from the 7.9, 11.9, 12.9 and 20
    $\mu$m observations (see text). The central point source,
    indicated by the '+', has a derived temperature of
    $\sim$~600~K}
  \label{fig:temp}
\end{figure}

We show the final images in Fig.~\ref{fig:nebula}~\&~\ref{fig:rings}.
Fig.~\ref{fig:temp} shows a temperature map, derived assuming that the
total flux in the images is equal to that seen in the ISO-SWS spectrum
\citepalias{iso_nature}. We realize that the variability of $\eta$~Car
may introduce errors in this calibration.  However, we verified the
calibration at 7.9 and 11.9 $\mu$m using $\gamma$~Cru and found values
of 1.6\,10$^4$ and 5.2\,10$^4$ Jy. The agreement between the two
methods is within 10$-$20 per cent. We stress that unless the shape of
the spectrum of $\eta$~Car has changed considerably between 1996 and
2001, our method should result in reasonable estimates of the
temperature. Note that the strong 10 $\mu$m silicate band
\citepalias[e.g.][]{iso_nature} may introduce some errors ($\sim$15\%)
in the temperature map.

\section{Description of the images}
The mid-IR image presented in Fig.~\ref{fig:nebula} shows the familiar
homunculus with an overall shape which is similar to that reported by
previous authors (e.g. \citealt{morse_hst,smith_98};
\citetalias{iso_nature,pantin_timmi}). At the core of the homunculus
we find a central point-like source which dominates the emission from
the nebula at short wavelengths. This point source very likely
coincides with the central star or binary. \citetalias{iso_nature}
show that the 2$-$8 $\mu$m wavelength range of the ISO spectrum of
$\eta$~Car is well represented by a power law, with a steeply rising
spectrum towards longer wavelengths. It is therefore reasonable to
assume that the central point source is responsible for the power law
component.  At 20 $\mu$m the central point source no longer dominates
the emission, which shows that there is not a large amount of cold
dust at this location. If the continuum in the 2$-$8 $\mu$m region is
due to thermal emission from dust, the power law nature of the
spectrum points to a flat temperature gradient in the hottest dust
(nearest to the central object), which is typically seen in optically
thick disks.  \emph{We stress that imaging at the sub-arcsecond scale
  is required to determine the true geometry of the innermost regions
  of the $\eta$~Car nebula.} Optical speckle imaging polarimetry
supports the presence of a disk \citep{falcke_speckle}.

The N band and Q band images give a detailed view of the innermost few
arcsec of the homunculus, where in the optical images the waist is
situated. Previous studies revealed the presence of emission blobs of
different intensity roughly 1.5 arcsec northeast and southwest of the
central point source. These emission blobs have been interpreted as
evidence for an equatorial torus \citepalias{polomski_99,iso_nature}.
This torus has been held responsible for the strongly bipolar geometry
of the homunculus \citepalias{iso_nature}. From the ISO spectrum a
temperature of the matter in the torus of 110 to 130~K was derived,
and a dust mass of about 0.15 M$_{\odot}$.

Our new images show for the first time the detailed geometry of the
material emitting at these wavelengths down to very low brightness
levels. The images show that the mid-IR emission is not due to
limb-brightening of a torodial dust distribution seen edge-on (in
\citetalias{iso_nature} assumed to be co-spatial with the distribution
of the massive cold material). Rather, the blobs reported by previous
studies turn out to be two arcs of emission. A careful inspection of
the 11.9 $\mu$m image shows that the arc southwest of the central star
is a closed ring. We will refer to these two structures as ``the two
rings'' in what follows, because both arcs have the same size and
inclination. The southwest ring passes through the central point
source, so this point source cannot be at the centre of the rings. The
northeast structure has an irregular surface intensity with a strong
intensity maximum in the north.  The axis connecting both rings is not
aligned with the projection of the long axis of the homunculus on the
sky (see below).  We note that one of the two rings can be recognised
in the 20 $\mu$m images published by
\citetalias{polomski_99,iso_nature} and \citetalias{pantin_timmi}.
However, the factor of two better spatial resolution at 10 $\mu$m
compared to 20 $\mu$m together with the high sensitivity allows for a
much clearer view of these structures.

The temperature map shows that the two rings have roughly similar
temperatures of 280$-$380~K. These temperatures agree well with the
values derived by previous studies (\citealt{smith_98};
\citetalias{polomski_99,pantin_timmi}). If we assume that there is no
large difference in the foreground extinction towards both rings, this
shows that the dust in the two rings is heated similarly.  The
question arises what the location of the 110-130~K massive dust
component, inferred from the ISO observations is. Since this spectral
component peaks at 30 $\mu$m and no flux jumps due to SWS aperture
transitions are seen, it must fit within the SWS band 3A beam. The low
temperature implies a different physical component from that in the
lobes or rings, such as a torus, as previously suggested by
\citetalias{iso_nature}.  \citet{2000Natur.405..532D} have shown that
this cold component should have a minumum projected area of 37
arcsec$^2$ in order to reproduce the required flux levels. An inclined
torus of that size (which would not show limb-brightening) fits easily
within the SWS beam.

\begin{figure}[t]
  \hbox{
    \psfig{figure=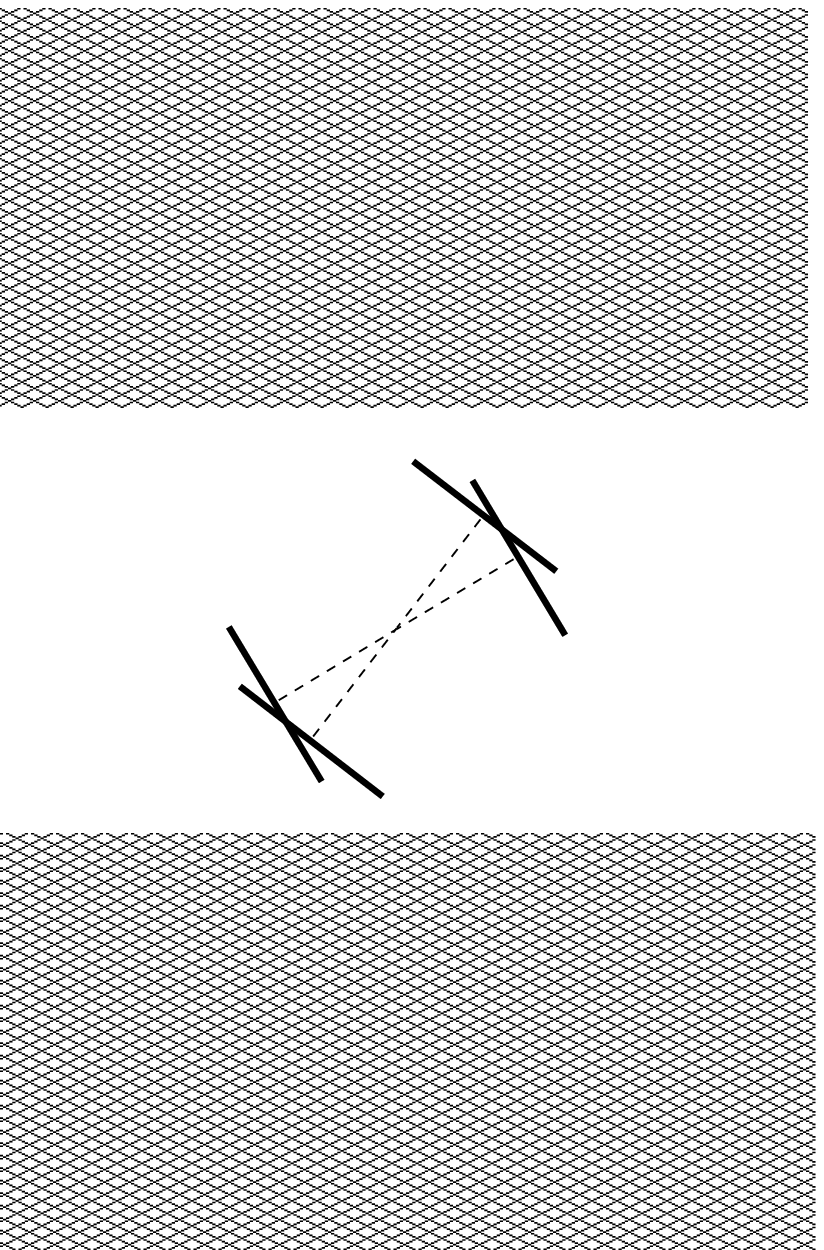,height=4.3cm}
    \hspace{0.5cm}
    \psfig{figure=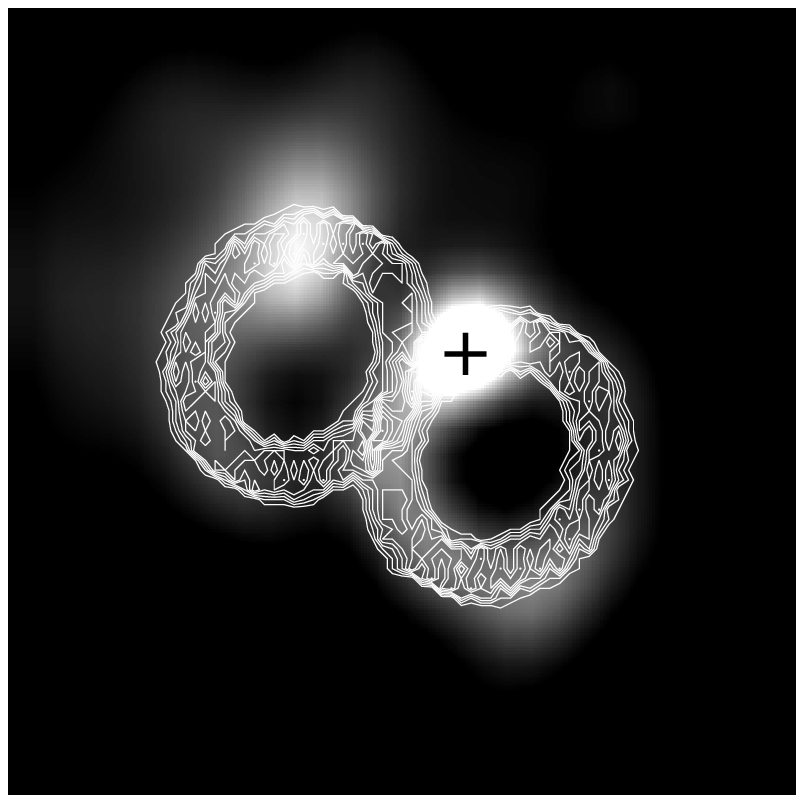,width=4.3cm}
    }
  \caption{Simple geometrical model.  The left panel sketches the
      position of the rings in a frame defined by the main axis of the
      homunculus and the axis connecting the rings.  There are two
      solutions because in the projection seen from Earth it is
      unknown which ring is in front.  The right panel shows the
    projected rings overlaid on the 11.9 $\mu$m image.  }
  \label{fig:geometry}
\end{figure}

We have constructed a geometrical model for the mid-IR emission of the
equatorial regions (sketched in Fig.~\ref{fig:geometry}).  The model
assumes the presence of two circular rings.  We varied the position
angle on the sky as well as the inclination angle until a good match
with the observations was obtained.  The result is shown in
Fig.~\ref{fig:geometry}.  We find that the rings have a diameter of
1.8 arcsec and a de-projected distance between the two rings of 3
arcsec.  The structure is rotated by 117 or 297 degrees in the plane
of the sky with respect to north-south (depending on which of the two
rings is in front).  The inclination between the major axis of the
homunculus and of the axis connecting the centres of the two rings is
either 37 or 58 degrees, using an inclination of the axis of the
homunculus to the plane of the sky of 40 degrees \footnote{The
  angle($\delta$) between the major axis of the homunculus($H$) and
  the axis connecting the centres of the rings($R$) is: $\cos\delta =
  (\sin\theta_H\sin\theta_R+\cos\theta_H\cos\theta_R)\sin i_H\sin
  i_R+\cos i_H\cos i_R$, where $i$ is the inclination and $\theta$ the
  position angle.}  .  The error on these angles is about 10 to 15
degrees, mostly determined by the wide range of values given in the
literature for the inclination of the homunculus
\citep[e.g.][]{hillier_etaneb92}.  Note that the region in our model
where the projection of both rings overlap on the sky coincides with
an intensity maximum in the observed images, again supporting our two
ring model.

\section{Discussion}
The new mid-IR images reveal a more complicated structure of the inner
few arcsec of the homunculus' equatorial regions than previously
suggested. The images show that two highly symmetric ring-like
structures are present, as well as an intensity maximum towards the
northeast. We find that the axis connecting the centre of the two
rings is inclined with respect to the major axis of the homunculus by
either 37 or 58 degrees, depending on which of the two rings is in
front, and assuming an inclination for the homunculus with respect to
the plane of the sky of 40 degrees. Kinematics of the gas in the rings
is needed to decide which of these two possibilities is correct.

\citetalias{pantin_timmi} show that the inner homunculus structure
seen at 20 $\mu$m represents regions of increased column density. If
indeed the rings trace a density enhancement, then they could be
denser rings in a bipolar nebula, similar to the rings in SN1987A
\citep{burrows_1987a_hst}.  We also note that this geometry shows
strong resemblance to that seen in PNe such as \object{He2-113}
\citep{sahai_he} and \object{Hb 12} \citep{1999ApJ...522L..69W}
including the two rings, the misalignment between the bipolar
structure and the rings, and the offset of the central star with
respect to the ring structure.  It seems reasonable to conclude that
the physical mechanism causing these structures is generic and acts in
high mass as well as in low mass objects.

A number of mechanisms to produce double rings in bipolar nebulae have
been proposed for PNe \citep{icke1988} and
SN1987A\citep{crotts_sn1987a}, but which one applies where has not
been established . In all cases the rings are perpendicular to the
major axis of the nebula, which for $\eta$~Car implies that the major
axis of this inner nebula is at a significant angle (37 or 58 degrees)
to the major axis of the homunculus.

It seems difficult to avoid the conclusion that there must have been a
change in the orientation of the outflow between the moment of
production of the homunculus and the creation of the double ringed
structure. This strongly favours the binary model for the $\eta$~Car
system. The shredded appearance of the skirt in the HST images and the
proper motion of the condensations indicate that the equatorial
regions were highly perturbed by the great eruption. It is therefore
likely that the rings were produced \emph{after} the great eruption.
1) The change of orientation could result from an asymmetry in the
mass loss during the great eruption.  2) It could be due to tidal
interaction of the eccentric binary with material in its environment.
The required mass for such a process can be estimated in the following
crude way. A gravitational perturbation can act most easily in the
apocenter. In a Keplerian motion about a mass $M_{*}$ with
eccentricity $e$ and semi-major axis $a$, the apocenter distance and
velocity are given by $r=a(1+e)$ and $v^2=\frac{GM_{*}(1-e)}{a(1+e)}$.
The required acceleration to change the orbital inclination by about 1
radian is of the order of $v^2/a(1+e)$.  Since the great eruption of
1840, about $N=160/5.5 \approx 29$ orbital periods have passed.  In
order to produce the required total change in $N$ steps, a disturbing
mass $\tilde{M}$ at distance R would have to fulfil the condition
$\frac{\tilde{M}}{M_{*}}(\frac{a}{R})^2 =
\frac{1}{N}\frac{1-e}{(1+e)^2} \approx 0.0054$, with $M_{*}$ the
reduced mass of the binary. Therefore with $e=0.6$ and $R\approx a$,
then $\tilde{M} \approx 5~10^{-3} \,M_{*}$.  A moderate amount of mass
close to the binary could already be sufficient to explain the
observed change in the system orientation.

\acknowledgements{We thank N. Ageorges for the excellent help with
  data-acquisition. VI thanks A. van Genderen for discussions. This
  work was supported by a NWO \emph{Pionier} grant to
  LBFMW and a NWO \emph{Spinoza} grant to E.P.J. van den Heuvel.
}

\bibliographystyle{apj}
\bibliography{eta}
\end{document}